\documentclass[bibyear]{aa}
\usepackage{graphicx}
\usepackage{scalerel}
\usepackage{txfonts}
\usepackage{amsmath}
\usepackage{amssymb}
\usepackage{float}
\usepackage{subcaption}
\usepackage{enumerate}
\usepackage{mathtools}
\usepackage{natbib}
\usepackage{hyperref}

\usepackage{xcolor}

\begin{document} 

    \title{Striving towards robust phase diversity on-sky}
    \subtitle{Implementing LIFT for VLT/MUSE-NFM}
    
    \author{
        Arseniy Kuznetsov \inst{1,2,3}
        \and
        Sylvain Oberti \inst{1}
        \and
        Benoit Neichel \inst{2}
        \and
        Thierry Fusco \inst{2,3}
        }
    \institute{
        European Southern Observatory, Karl-Schwarzschild-str-2, 85748 Garching, Germany\\
        \email{akuznets@eso.org}
        \and
        Aix Marseille Univ, CNRS, CNES, LAM, 13013 Marseille, France
        \and
        DOTA, ONERA, Universit\'e Paris Saclay, F-91123 Palaiseau, France
        }
    
    \date{Received XXXXX XX, XXXX; accepted XXXXX XX, XXXX}

\abstract
   {The recent IRLOS upgrade for VLT/MUSE narrow field mode (NFM) introduced a full-pupil mode to enhance sensitivity and sky coverage. This involved replacing the 2$\times$2 Shack-Hartmann sensor with a single lens for full-aperture photon collection, which also enabled the engagement of the linearized focal-plane technique (LIFT) wavefront sensor instead. However, initial on-sky LIFT experiments have highlighted a complex point spread function (PSF) structure due to strong and polychromatic non-common path aberrations (NCPAs), complicating the accurate retrieval of tip-tilt and focus using LIFT.}
   {This study aims to conduct the first on-sky validation of LIFT on VLT/UT4, outline challenges encountered during the tests, and propose solutions for increasing the robustness of LIFT in on-sky operations.}
   {We developed a two-stage approach to focal-plane wavefront sensing, where tip-tilt and focus retrieval done with LIFT is preceded by the NCPA calibration step. The resulting NCPA estimate is subsequently used by LIFT. To perform the calibration, we proposed a method capable of retrieving the information about NCPAs directly from on-sky focal-plane PSFs.}
   {We verified the efficacy of this approach in simulated and on-sky tests. Our results demonstrate that adopting the two-stage approach has led to a significant improvement in the accuracy of the defocus estimation performed by LIFT, even under challenging low-flux conditions.}
   {The efficacy of LIFT as a slow and truth focus sensor in practical scenarios has been demonstrated. However, integrating NCPA calibration with LIFT is essential to verifying its practical application in the real system. Additionally, the proposed calibration step can serve as an independent and minimally invasive approach to evaluate NCPA on-sky.}


   \maketitle

\section{Introduction}
\label{sec:intro}

Adaptive optics (AO) represents a significant advancement in the field of optical imaging, enabling the correction of wavefront distortions that occur as light passes through turbulent media or imperfect optical systems. The wavefront is corrected using deformable mirrors (DM) based on measurements performed by wavefront sensors (WFSs).

In astronomical AO systems, incoming light is usually divided into distinct optical paths: one fraction is channelled towards scientific instruments (science path), while another fraction is directed to WFSs (AO path). However, this scheme introduces a possibility for path-specific aberrations termed NCPAs \citep[see][]{Mugnier:08, Lee:97}. For instance, aberrations that occur outside the AO optical path remain unseen by WFSs and, thus, they cannot be compensated for. Conversely, aberrations that come from WFSs are not inherent to the scientific path but can still affect the science image being propagated through the AO feedback loop. The uncalibrated NCPAs can prominently deteriorate the quality of the resulting scientific PSFs, preventing the AO system from delivering a high Strehl ratio (SR). This underscores the importance of measuring and compensating for NCPAs.

In this framework, leveraging the focal-plane point spread function (PSF) for wavefront sensing (WFSing) and phase retrieval emerges as an appealing option. This approach facilitates extracting information directly from post-AO PSFs obtained close to the science path. As a result, it allows for the measurement of specific NCPAs that impact the quality of scientific images.

In addition, post-AO PSFs imprint the information about phase discontinuities \citep{Lamb:17}. These effects are typically associated with co-phasing errors \citet{Chanan:98} or with the low wind effect (LWE) \citet{Sauvage:16, Pourre:22}. In turn, the latter can introduce wavefront errors with root-mean-square (RMS) wavefront error (WFE) reaching hundreds of nanometers; this is manifested as combinations of the differential piston and tip-tilt errors that significantly deteriorate the image quality. It is important to note that this effect will be amplified on the upcoming extremely large telescopes, making the question of its measurement increasingly important. Currently, multiple methodologies have been proposed to address petaling (island) effects \citep{Esposito:03, Shi:04}. Among them, focal plane-based techniques can also be applied \citep[see][]{Agapito:22, Wilby:18}.

To summarise, leveraging phase retrieval in the focal plane proves to be helpful for measuring LWE, co-phasing segmented telescopes \citep{Lamb:21, Acton:12}, characterising NCPAs in AO systems \citep{Lamb:18, Robert:08}, and image post-processing \citep{Wilding:17, Wu:18}. Focal-plane WFSing can successfully complement other WFSing approaches and thus unlock more nuanced AO control strategies, while also enabling more detailed measurements of optical wavefronts. Moreover, depending on the specific AO system, PSFs can be abundant, appearing not only in wavefront sensor images but also in the science images themselves.
We highlight a few of the existing focal-plane phase retrieval techniques in the section below.

\section{A concise overview of focal-plane WFSing techniques}
\label{sec:overview}

\subsection{Phase diversity (PD)}
Phase diversity is a versatile family of focal-plane techniques used for WFSing. It involves capturing images at single or multiple focal planes, each corresponding to a known phase diversity introduced into the optical path. This diversity helps to resolve sign ambiguity and optimises the distribution of information about PSF and object morphologies across the detector pixels. While the classical implementation from~\citet{Gonsalves:82} employs defocus diversity, other aberrations and types of diversities can also be used \citep{Campbell:04, Brady:09, Almoro:08}.

Among PD techniques, the estimation of the phase in the pupil plane usually turns into a problem of finding the optimal set of modal coefficients that minimises the error between the model and data. In this approach, expressing phase as a superposition of pupil-space modes entails implicit regularisation, increasing robustness to noise. However, a solution for the phase in the pixel space is also possible, as outlined in \citet{Robert:08}. The solution can be found by iteratively fitting the PSF model to the data. It usually involves first or second-order \citep{Smith:13} approximations of the PSF model in the vicinity of phase estimates. Meanwhile, having multiple diverse images also allows for the decoupling of the unknown object morphology from wavefront aberrations \citep{Mocoeur:09, Blanc:03}.
Overall, PD is a valuable tool for measuring wavefront aberrations and improving image quality. In AO, PD methods are commonly employed due to their minimally invasive implementation and strong performance in the presence of noise. It can effectively handle a wide range of aberrations, making it a resilient and versatile technique for focal-plane WFSing.

\subsection{Gerchberg-Saxton-based (GS) algorithm}
The GS algorithm described in \citet{Gerchberg:72}, \citep[see also][]{Fienup:82, Huang:21} is a classical forward-reverse phase retrieval algorithm that has given rise to such techniques as PD. It alternates between the spatial and Fourier domains, enforcing the amplitude with known measurements while updating the phase. The process iterates until the phase information converges, providing a phase estimate consistent with amplitude constraints applied in both domains.
It usually requires many steps to converge and may easily become trapped in local minima due to the vast space of parameters to optimise. Therefore, it does not guarantee global convergence to the true phase, making this method unsuitable for real-time applications.

\subsection{Linearized focal-plane technique (LIFT)}
LIFT is a PD-based technique that employs a known amount of astigmatism instead of defocus to create phase diversity \citep[see][]{Meimon:10, Plantet:14, Plantet:17}. Consequently, there is no need to employ additional extra-focal intensity planes to resolve the sign ambiguity, meaning that only one image can be used for phase retrieval. This greatly simplifies the hardware implementation of LIFT and prevents flux splitting between images, leading to even greater sensitivity. However, it also makes it difficult to decouple the object morphology from wavefront aberrations. To summarise, LIFT is an ideal choice for low-order mode estimation using a point source in low-flux conditions.

\subsection{Fast \& furious (F$\&$F)}
Fast \& Furious is a very performant PD-based method designed to run in closed-loop conditions \citep{Korkiakoski:14, Bos:20}. It is essentially derived from the small-phase approximation outlined in \citet{Gonsalves:01} and it employs the difference between the DM shapes on current and preceding loop iterations as phase diversity. It also uses focal-plane PSFs acquired on the current and previous iterations to reconstruct the wavefront. Unlike LIFT, F$\&$F does not disturb the morphology of the in-focus PSF. However, it also requires persistent access to the DM commands to extract the diversity phase.

\subsection{Machine learning (ML) methods}
Recently, ML-based methodologies have gained popularity due to their flexibility and versatility. In contrast to conventional methods, ML approaches can effectively integrate complex non-linear relationships without relying on explicit analytical models \citep[see][]{Lloyd:92, Wang:21, Terreri:22}. Therefore, they can receive inputs from various sources and be applied to virtually any kind of AO system. Such models can be trained both offline and online (in-loop) on actual data, enabling them to adjust to evolving conditions or changing system configurations. Nonetheless, in data-driven approaches, estimation accuracy directly depends on the quality and size of the training dataset. Despite the abundance of PSFs in AO systems, collecting a large and diverse, real or artificial dataset can prove challenging in practice. Moreover, running ML solutions in-loop necessitates significant computational overheads. In addition, the opaque nature of ML-based approaches may be viewed as a disadvantage in certain applications. Nevertheless, ML presents promising prospects in many applications, including focal-plane WFSing.

\vspace{20pt}
Selecting the appropriate method for focal-plane WFSing depends on the specific system or application in question. In the context of this work, our primary focus is the InfraRed Low-Order wavefront Sensor (IRLOS). It is an integral component of the Adaptive Optics Facility (AOF) \citep[see][]{Oberti:18}, which enables the narrow field mode (NFM) of Multi Unit Spectroscopic Explorer (MUSE) instrument. The primary goal of IRLOS is to accurately sense low-order (LO) modes, specifically tip and tilt (TT) and defocus. This capability is crucial as TT modes are beyond the sensing capability of laser guide stars (LGSs), whereas defocus retrieval is compromised by the altitudinal drift of the sodium layer.
To enable tip-tilt and focus (TTF) measurements, IRLOS employs a single natural guide star (NGS). The baseline configuration of IRLOS is a 2$\times$2 Shack-Hartmann (SH) WFS, operating in J and H optical bands, with a current limiting NGS magnitude of $m_J < 19$. 

A proposal has been made to enhance the sensitivity of IRLOS by substituting the current SH WFS configuration with a full-pupil mode. The full-pupil mode entails the removal of the 2$\times$2 lenslet array from the pupil plane of IRLOS and replacing it with a single lens, thereby eliminating the per-aperture flux splitting and enabling the collection of photons from the entire pupil and focusing them into a single PSF. In this regime, IRLOS would operate as a fast TT sensor at the frequency of 200 Hz or 500 Hz (depending on the magnitude of a target) and as a slow truth sensor for defocus, accumulating a PSF over 1-2 seconds of exposure. Due to the broadband sensitivity of IRLOS and the necessity to accumulate as much flux as possible under low-flux conditions, PSFs recorded in the full-pupil mode are polychromatic.

In this context, LIFT was chosen as the optimal slow defocus sensing method due to its inherent noise robustness, ease of hardware implementation, and the ability to accurately estimate defocus using a single image in a single focal plane. Furthermore, a slow defocus loop operates at a frequency of 0.5-1 Hz, providing LIFT with sufficient time to perform computations.
The question of sensitivity gains when using LIFT compared to the 2$\times$2 SH WFS is covered in detail in \citet{Plantet:14} and \citet{Meimon:10}. However, the main objective of this paper is to discuss the challenges faced while applying LIFT on-sky and suggest methods to enhance its robustness. However, it is useful to briefly investigate the theoretical foundations of this method first to provide a background for later discussions.

\section{Theory}
\label{sec:theory}

The focal-plane PSF with introduced phase diversity can be approximated using the following expression:

\begin{equation}
   \begin{aligned}
       \mathbf{I}(\mathbf{r}, \Phi(\mathbf{u})) = \mathbf{O}(\mathbf{r}) \circledast 
       \sum_{\lambda} F(\lambda) \left| \hspace{2pt} \mathcal{F} \{ \mathbf{P}(\mathbf{u}) \hspace{1pt} e^{ i\frac{2\pi}{\lambda} \mathbf{\Phi}(\mathbf{u}) } \} \hspace{1pt} \right|^2 \hspace{-3pt} \Delta\lambda \hspace{2pt} + \hspace{2pt} \mathbf{n}(\mathbf{r}),
   \end{aligned}
   \label{eq:theory:I_A}
\end{equation}
\noindent
where $\mathbf{P}(\mathbf{u})$ is the aperture mask, $\mathbf{\Phi}(\mathbf{u})$ is the optical path difference (OPD) in the pupil and $\mathbf{r}$, $\mathbf{u}$ are the 2D pixel coordinates in the focal and pupil planes, respectively. In subsequent expressions, we omitted the pixel coordinates. Also, $\mathbf{n}(\mathbf{r})$ is the per-pixel noise term in image space, $\mathbf{O}(\mathbf{r})$ is the convolution kernel that represents the shape of an NGS in case it is an extended source, $\circledast$ is the 2D convolution operation, and $\mathcal{F}\{ \cdot \}$ is the Fourier transform, while $\lambda$ is the wavelength. Quantities in bold are the vectors and matrices. Then, $\mathbf{\Phi}(\mathbf{u})$ can be described as the superposition of modes in the pupil space:

\begin{equation}
    \begin{aligned}
        \mathbf{\Phi} = \sum_{i}{a_i \mathbf{Z}_i} + \mathbf{\phi}_d = \mathbf{Z}\mathbf{A}+\mathbf{\phi}_d 
    \end{aligned}
    \label{eq:theory:OPD}
,\end{equation}

\noindent
where $\mathbf{Z}_i (\mathbf{u})$ is the $i$-th pupil mode of modal basis $\mathbf{Z}$. In this work, we adopt the Zernike modal basis. However, the choice of the basis can be arbitrary under the condition of orthogonality in the pupil plane. The vector $\mathbf{A} = [a_0, a_1, \ldots, a_M]$ comprises the modal coefficients $a_i$, $\mathbf{\phi}_d (\mathbf{u})$ denotes the introduced diversity phase. We specifically employ astigmatic phase diversity using the mode $Z_5$\footnote{Following the indexing from \citet{Noll:76}.}. This choice is connected to the hardware implementation of LIFT, details of which will be expanded in the next section. However, any predetermined wavefront aberration that avoids sign ambiguity and significantly impacts the PSF morphology can function as diversity. The question of choosing the optimal amplitude of phase diversity is covered in \citet{Plantet:13}, \citet{Dean:03}, and \citet{Polo:13}.

If $P(\mathbf{u})$, $Z_i(\mathbf{u})$, and $\mathbf{O}(\mathbf{r})$ are pre-defined, the PSF morphology described using Eq. (\ref{eq:theory:I_A}) is a function of modal coefficients $\mathbf{A}$; thus, we can say that $\mathbf{I}(\mathbf{r}, \Phi(\mathbf{u})) = \mathbf{I}(\mathbf{A})$. Therefore, the goal of the focal-plane wavefront estimator is to find the optimal vector of modal coefficients $\hat{\mathbf{A}}$ that minimises the difference  $\Delta \mathbf{I} (\mathbf{A})$ between the observed PSF $\mathbf{I}_\text{data}(\mathbf{r})$ and the simulated PSF $\mathbf{I}(\mathbf{A})$, as:

\begin{equation}
    \Delta \mathbf{I} (\mathbf{A}) = \Delta \mathbf{I} = \mathbf{I}_\mathrm{\hspace{1pt} data} - \mathbf{I}(\mathbf{A}).
\end{equation}
\noindent
We can simplify further derivations in this section by omitting the convolution kernel $\mathbf{O}(\mathbf{r})$ and assuming a PSF at a single wavelength.

The derivation of the LIFT estimator is covered in detail in \cite{Meimon:10} and \citet[][chap.~A.2]{Plantet:these}. Therefore, we do not delve into the derivations here. The best estimate $\hat{\mathbf{A}}$ for the modal coefficients can be derived within the maximum likelihood (ML) or maximum a posteriori probability (MAP) frameworks (Eqs. (\ref{eq:theory:A_ML}) and (\ref{eq:theory:A_MAP}), respectively) as:

\begin{eqnarray}
    \hat{\mathbf{A}}_\text{ML}  & = & \left( \mathbf{H}^\top \mathbf{R}^{-1}_\mathrm{n} \mathbf{H} \right)^{-1} \hspace{3pt} \mathbf{H}^\top \mathbf{R}^{-1}_\mathrm{n} \Delta \mathbf{I} \label{eq:theory:A_ML}, \\
    \hat{\mathbf{A}}_\text{MAP} & = & \left( \mathbf{H}^\top \mathbf{R}^{-1}_\mathrm{n} \mathbf{H} + \mathbf{C}_\varphi^{-1} \right)^{-1} \hspace{3pt} \left( \mathbf{H}^\top \mathbf{R}^{-1}_\mathrm{n} \Delta \mathbf{I} + \mathbf{C}_\varphi^{-1} \bar{\mathbf{A}} \right). \label{eq:theory:A_MAP}
\end{eqnarray}

\noindent
Here, $\mathbf{H} = \mathbf{H}(\mathbf{A}) = \left[\frac{\partial \mathbf{I}(\mathbf{A})} {\partial a_0} \cdots \frac{\partial \mathbf{I}(\mathbf{A})}{\partial\ a_M} \right]$ is the interaction matrix (Jacobian), $\mathbf{R}_\mathrm{n} = \langle \mathbf{n} \mathbf{n}^\top \rangle$ is the covariance matrix of detector pixels, assuming they are uncorrelated with each other and that the distribution of photo-electrons per pixel is Gaussian.

Eqs. (\ref{eq:theory:A_ML}) and (\ref{eq:theory:A_MAP}) provide the one-shot estimations derived by linearising the $\Delta \mathbf{I}(\mathbf{A})$ in the vicinity of $\mathbf{A} = \left[ 0, 0, \ldots, 0 \right]$, resulting in a very limited linearity range. To overcome this limitation, the ML and MAP estimators can be approached iteratively when every subsequent estimate $\mathbf{\hat{A}}_{i+1}$ is computed in the vicinity of the previous estimate $\mathbf{\hat{A}}_{i}$ as follows:

\begin{eqnarray}
    \hat{\mathbf{A}}^\text{ML}_{i+1} & = & \left( \mathbf{H}(\hat{\mathbf{A}}_i)^\top \mathbf{R}^{-1}_\mathrm{n} \mathbf{H}(\hat{\mathbf{A}}_i) \right)^{-1} \nonumber \\
    & & \mathbf{H}(\hat{\mathbf{A}}_i)^\top \mathbf{R}^{-1}_\mathrm{n} \Delta \mathbf{I}(\hat{\mathbf{A}}_i) + \hat{\mathbf{A}}_i \label{eq:theory:iterative_ML}, \\
    \hat{\mathbf{A}}^\text{MAP}_{i+1} & = & \left( \mathbf{H}(\hat{\mathbf{A}}_i)^\top \mathbf{R}^{-1}_\mathrm{n} \mathbf{H}(\hat{\mathbf{A}}_i) + \mathbf{C}_\varphi^{-1} \right)^{-1} \nonumber \\
    & & \left( \mathbf{H}(\hat{\mathbf{A}}_i)^\top \mathbf{R}^{-1}_\mathrm{n} \Delta \mathbf{I}(\hat{\mathbf{A}}_i) + \mathbf{C}_\varphi^{-1} \bar{\mathbf{A}} \right) + \hat{\mathbf{A}}_i \label{eq:theory:iterative_MAP}  
.\end{eqnarray}
\noindent
Here, $\mathbf{C}_\varphi^{-1}$ is the covariance matrix of the modal coefficients, and $\bar{\mathbf{A}}$ is their mean.
However, determining the exact statistics of the modal coefficients is challenging since LIFT operates in closed-loop conditions, and very few on-sky measurements have been performed so far. At the same time, the imprecise knowledge of coefficient statistics can lead to biased solutions. Therefore, within the scope of this work, we computed $\hat{\mathbf{A}}$ using an unbiased iterative ML estimator.
However, in practice, it is possible to assume that $\mathbf{C}_\varphi^{-1}$ follows some power law while $\bar{\mathbf{A}} = \left[ 0, 0, \ldots, 0\right]$ when utilising MAP estimation framework.

In essence, the iterative estimators shown in Eqs. (\ref{eq:theory:iterative_ML}) and (\ref{eq:theory:iterative_MAP}) are similar to the Gauss-Newton optimisation method \citep[see][chap.~9]{Ake:96}, while the minimisation criterion is equivalent to the mean squared error (MSE) weighted by $\mathbf{R}^{-1}_\mathrm{n}$. From this point of view, LIFT is a least-squares minimisation problem of fitting the PSF model to the data that can be decomposed into three main components: PSF model, non-linear solver (optimisation method), and optimisation criterion (loss function).
Depending on the required application, each component can be replaced or adjusted in this context. This provides an essential framework for the further enhancement of LIFT. For example, the PSF model can be modified. Furthermore, the Gauss-Newton optimiser can be substituted with other approaches, such as the L-BFGS method described \citet{Liu:89}. Such modifications also suggest a notable deviation from the expressions in Eqs. (\ref{eq:theory:iterative_ML}) and (\ref{eq:theory:iterative_MAP}). The issue of modifying LIFT is further examined in Sect.~\ref{subsec:DIP}. The upcoming section will discuss the practical application of the unaltered LIFT technique on real-life sky data.

\section{First on-sky tests}
\label{sec:onsky_1}

Following the mid-2021 upgrade of IRLOS, the feasibility of LIFT was experimentally tested in on-sky conditions on Unit Telescope 4 (UT4) of ESO's Very Large Telescope (VLT) facility. In the hardware implementation of LIFT within IRLOS, a 2$\times$2 lenslet array was replaced with a single lens and a cylindrical lens was added to induce astigmatism in the optical path. This lens determines the amplitude of introduced astigmatic diversity, resulting in roughly -170 nm RMS.
However, since the astigmatic lens occupies a slot in the filter wheel, it cannot be used simultaneously with a narrow-band filter. As a result, LIFT needs to operate with strongly polychromatic PSFs that cover both J and H bands.
Additionally, the SAPHIRA detector \citep[see][]{Finger:14}. was introduced with the IRLOS upgrade. It yielded an angular resolution of $\sim$9.7 mas per pixel in full-pupil mode\footnote{With the LIFT lens introduced.}, resulting in a sampling of $\sim$3.2 and $\sim$4.4 pixels per $\frac{\lambda}{D}$ at the J and H bands, respectively. Consequently, LIFT has access to PSFs that are above the Nyquist sampling and have well-resolved morphological features.

\begin{figure}[H]
   \centering
   \vstretch{1.0}{ \includegraphics[width=0.475\textwidth]{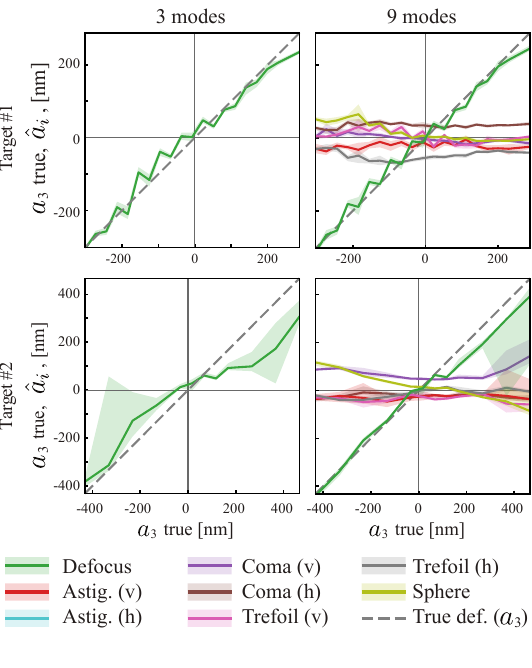} }
   \caption{
   LIFT estimations for the dataset of PSFs with ramping defocus. Results were obtained for two on-sky targets (top and bottom rows). Estimations were done for three modes $ ( a_1, a_2, a_3 ) $ (left column) and nine modes $( a_1, \ldots a_{10} )\setminus \{ a_5 \}$ (right column). For target \#2, including modes beyond TTF allowed us to retrieve the defocus in an accurate way. Here, TT modes are estimated but not displayed. Lines represent median values, while error bars show $1\sigma$ confidence interval.
   }
   \label{fig:exp:onsky1}
\end{figure}

The first experimental validation of LIFT was conducted on two bright natural targets (see Fig.~\ref{fig:exp:onsky1}). These tests aimed to introduce the range of known defocus values and retrieve them utilising LIFT. Defocus was introduced by shifting the focusing stage of MUSE. Therefore, the introduced defocus offsets are the only accessible ground truth values in this experiment.
The selected range of defocus values was based on simulations which showed that the linearity of LIFT in the H-band is confined to about $\pm$410 nm RMS\footnote{Assuming that the initial guess for defocus is zero.}.
Since defocus was the sole variable changed during the experiment, other coefficients are expected to exhibit relative stability. Table~\ref{tab:onsky:params_1} lists the exact data acquisition parameters.

\begin{table}[hbt!]
   \caption{Acquisition parameters for the first on-sky test.}
   \label{tab:onsky:params_1}
   \centering
   \begin{tabular}{lc}
       \hline
       \noalign{\smallskip}
       Parameter & Values \\
       \noalign{\smallskip}
       \hline
       \noalign{\smallskip}
       $m_J$ of NGS targets & 9, 12.5 \\
       Exposure time per PSF, [s] & 1 \\
       Seeing, ["] & 0.5, 1.5 \\
       PSFs recorded per defocus offset & 10, 20 \\
       Introduced defocus, [nm RMS] & -432 \ldots 467, -302 \ldots 288 \\
       $a_5$ diversity (lens), [nm RMS] & -170 \\
       Target wavelength $\lambda$, [nm] & 1215 \ldots 1625 \\ 
       $\sigma_{obj}$, [mas] & 12, 14.5 \\
       \noalign{\smallskip}
       \hline
   \end{tabular}
\end{table}

The results presented in Fig.~\ref{fig:exp:onsky1} demonstrate the successful retrieval of defocus with LIFT alongside the values of additional coefficients. The saw-tooth behaviour seen in coefficient estimates for target \#1 is caused by the fact that every consequent PSF in the scan was recorded with alternating defocus sign, which required moving the focusing stage of MUSE to opposite directions in ping-pong fashion after every recorded sample, causing the introduction of the unmeasured bias into the reference values due to the hysteresis of the focal stage. The constant bias in defocus estimates for the second target is caused by the fact that two parts of the same scan were recorded in two separate runs, potentially biasing the reference.

\subsection{Lessons learned}
\label{sec:onsky_1:lessons}

Although initial on-sky experiments clearly demonstrated the capability of LIFT to operate in realistic conditions, achieving these results practically demanded multiple manual modifications to the estimation process. Therefore, discussing these adjustments is essential for the interpretation of the results.
\vspace{2pt}

First, LIFT required an initial approximation of the defocus. Providing LIFT with a defocus prior yielded faster and more accurate convergence, especially when dealing with strongly defocused PSFs. The initial guess for $a_3$ was extracted from the diversity-induced PSF elongation by fitting a 2D Gaussian to each PSF and computing the $A \hspace{1pt} (\sigma_x / \sigma_y - 1) + B$ ratio, where $\sigma_x, \sigma_y$ are the fitted parameters of a 2D Gaussian, while $A, B$ are the normalising constants for calibration.
\vspace{2pt}

Second, adding a convolution kernel $\sigma_{obj}$ to the PSF model was required. In the PSF model defined in Eq. (\ref{eq:theory:I_A}), the convolution kernel $\mathbf{O}(\mathbf{r})$ represents an extended object. However, for this experiment, we selected only point-source targets, meaning that this kernel absorbed all the effects that are not accounted for by the PSF model used in LIFT. The primary contributors may include the residual TT jitter, the polychromatic smearing of PSFs, and the potential cross-talk between the detector's pixels (although the latter effect is negligible in the SAPHIRA detector). The atmospheric dispersion can also have an impact on the IRLOS PSFs because the dispersion compensator situated upstream of IRLOS is designed to optimise wavelengths in the visible spectrum for MUSE and may produce residual elongation of the spot due to chromatism at J and H bands.
The initial shape of the 'object' can be derived from the expected TT jitter under specific seeing conditions and then adjusted via fitting. Then, $\mathbf{O}(\mathbf{r})$ is modelled as a 2D Gaussian. However, adding $\sigma_{obj}$ as an optimisable parameter is undesirable due to the single-image nature of LIFT, as it can cause coupling between the kernel parameters and modal coefficients. Therefore, we had to fine-tune the convolution kernel parameters manually.
\vspace{2pt}

\begin{figure}[h!]
    \centering
    \vstretch{1.0}{ \includegraphics[width=0.475\textwidth]{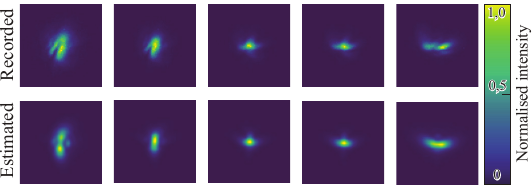} }
    \caption{Examples of recorded PSFs for target \#1 (top row) and their estimated counterparts (bottom row), captured with varying defocus levels and inserted diversity. The prominent effect of NCPAs on PSF morphology is evident. The data was acquired in the focal plane of IRLOS after closing the HO and fast TT control loops, with a seeing of $\approx$ 0.5".}
    \label{fig:exp:onsky1_PSFs}
 \end{figure}

Third, modes beyond focus had to be estimated. While target \#1 indicated that retrieving defocus by estimating only TTF is feasible, target \#2 showed that estimating additional modes is sometimes essential for accurate defocus retrieval. It was determined through experimentation that the most precise results were obtained by estimating the first ten Zernike modes (excluding $a_5$, the diversity astigmatism). Similarly, although LIFT was proposed for slow defocus sensing, incorporating TT modes into the estimation enhanced the accuracy of estimation by compensating for sub-pixel shifts of the PSF relative to the centre of the region of interest, even though TT estimates from LIFT are not used for control. Meanwhile, the necessity of estimating orders higher than TTF can be understood when examining the recorded on-sky PSFs (refer to Fig.~\ref{fig:exp:onsky1_PSFs}), which reveal highly complex morphological features. We attribute this to the strong presence of the uncompensated NCPAs. We further suggest that these NCPAs can be modelled as phase aberrations in the pupil plane. Additionally, their quasi-static behaviour is indicated by the temporal consistency of speckle patterns observed on PSF images and the relative stability of estimated coefficients, as demonstrated in Fig.~\ref{fig:exp:onsky1}.
Thus, including only TTF modes or even only focus in the estimation is generally insufficient for capturing this complexity. Adding more modes improves the descriptiveness of the model. However, it also introduces more variables, weakening the inherent regularisation induced by the modal decomposition of the pupil phase. This makes the model more susceptible to noise and overfitting, leading to cross-talk between modal coefficients. For example, the astigmatic diversity mode was excluded from the estimation due to its coupling with other modal coefficients. 
This scenario underscores a trade-off between complexity and precision. Reducing the number of estimated modes regularises the solution, yet a model that is too simplistic may result in unrepresented higher-order modes aliasing into the TTF. We offer additional information on cross-talk in Appendix~\ref{sec:appendix:overfit}.

Out of all the points mentioned, the last one needs the most attention. Estimating excessive modes restricts the robustness of the method, particularly in low-flux scenarios, as later demonstrated in Fig. \ref{fig:exp:onsky2_sensitivity}. This contradicts the initial promise of using LIFT as a highly sensitive TTF sensor. Addressing this issue is therefore critical.

\subsection{Need for LIFT calibration}
We propose addressing this problem by separating TTF retrieval from NCPA estimation. This can be achieved in a configuration divided into two consecutive steps. The first step is the model calibration. It 'absorbs the model complexity' by retrieving NCPAs, assuming their quasi-static behaviour. This step is done offline before closing the slow TTF loop with LIFT. The second step is the focus estimation (online), which estimates only TTF, leveraging the preceding NCPA calibrations as a prior.

Calibration for NCPA can be conducted using either an artificial source or directly an on-sky target. However, using the latter can yield more realistic results due to the ability to capture all contributors missed by the optical path of the artificial calibration source\footnote{At least, in the case of AOF.}. Our study thus focuses on developing a methodology that leverages on-sky observations for retrieving quasi-static NCPAs, utilising the same information channel as LIFT for minimal invasiveness and ease of execution. The next section will discuss further details on implementing this calibration approach.

It should be noted that within the scope of this paper, the term 'quasi-static' refers to the assumption that the measured NCPAs remain constant during the calibration process and continue to be applicable when the loop is closed on defocus. However, the detailed investigation of the NCPAs' exact temporal dynamics to determine how long the calibration estimates remain valid is not addressed in this paper.

\section{Two-stage approach}
\label{sec:2_stage}

The following section focuses on implementing and analysing the two-stage focal-plane WFSing technique. All the results presented in this section are derived from realistic and representative IRLOS simulations.

\subsection{The problem of NCPA calibration}
\label{sec:2_stage:calib}

We start the discussion with a more detailed description of the calibration stage. The primary objective and essential criteria are as follows. It must accurately estimate the quasi-static component of NCPAs and reconstruct as many modes as possible to provide enough degrees of freedom to describe the complex PSF morphology accurately. The cross-coupling between estimated modes should be minimised. Similar to LIFT, calibration should utilise PSF images acquired in the focal plane of IRLOS. Calibration must be executed after closing the high order (HO) loop and before performing in-loop focus measurements.
Calibration is conducted offline and thus is not subject to strict hardware and time constraints encountered during in-loop operations.

To accurately estimate wavefront aberrations and access higher-order modes, it is essential to retrieve high spatial frequencies in the pupil plane and thus preserve the structure of speckles. This requires recording monochromatic PSFs to eliminate the effects associated with polychromaticity. Additionally, using short exposure times can reduce the blurring associated with the uncompensated HO and TT residuals. Therefore, if a point source is used, $\mathbf{O}(\mathbf{r})$ can be excluded from the PSF model because the main contributors, such as TT jitter and polychromatic blurring, are mitigated in this case (see Sect.~\ref{sec:onsky_1:lessons}). Therefore, the PSF can now be modelled using solely wavefront aberrations. In this case, a bright target must be used, which is necessitated by the combination of short exposures combined with the narrow spectral bandwidth. These factors mean that the calibration and defocus-retrieval stages may utilise different on-sky targets. This also implies that the data acquisition parameters for the calibration stage must differ from those used for LIFT.

The IRLOS WFS utilises SPARTA \citep[see][]{Valles:12} as a real-time computer (RTC). In the current configuration, SPARTA allows for the recording and storing of the detector pixel readings from the SAPHIRA detector at a maximum frequency of 10 Hz. Consequently, it enables access to frames with 2-5 ms exposure time at 10 Hz or to averaged images with a 0.1 second exposure at 10 Hz. In terms of exposure duration, theoretically, an infinitely short exposure would yield the most accurate, unaveraged snapshot of speckle patterns, thereby giving access to the highest number of spatial frequencies. Conversely, our simulations have demonstrated that extending the average exposure beyond 0.15-0.2 seconds leads to a predominance of HO residuals. Therefore, within the framework of this study, we have selected this upper limit as a compromise to optimise the trade-off between minimising exposure time and maximising the signal-to-noise ratio (S/N).

\subsection{Different approaches to modal coefficients optimisation}
\label{subsec:methods}

Considering the requirements mentioned earlier in this paper and leveraging the hardware capabilities, it becomes possible to capture PSF cubes. For example, with a sampling rate of 10 Hz, capturing an entire PSF dataset within one minute of exposure time is feasible. This effectively increases the quantity of available data samples, better defining the problem of searching for optimal modal coefficients. Therefore, we propose leveraging the PSF cubes instead of averaged long-exposure PSFs for phase retrieval. In this scope, several non-linear optimisation approaches can be considered. Before elaborating on this, a few notations must be introduced in the table below.

\begin{table}[h!]
   \caption{Notations used in this and subsequent sections.}
   \label{tab:DIP:notations}
   \centering
   \begin{tabular}{ll}
       \hline
       \noalign{\smallskip}
       Notation & Meaning \\
       \noalign{\smallskip}
       \hline
       \noalign{\smallskip}
       N & number of PSFs in the cube, \\
       M & number of simulated or estimated modes, \\
       W, H & width and height of the PSF image, \\
       $\mathcal{I}_\mathrm{\hspace{-1pt} data}$  & cube of recorded PSFs, \\
       $\mathbf{I}_{\mathrm{\hspace{1pt} data}, \hspace{1pt} i}$ & $i$-th PSF in the cube (sample), $i = 1,\ldots N$, \\
   \end{tabular}
\end{table}

\begin{table}[h!]
    \centering
    \begin{tabular}{ll}
       $\mathbf{A}_i$ & $i$-th vector of modal coefficients, \\
       $\mathcal{A}$ & matrix of stacked coefficient vectors, \\
       $\mathbf{I}(\mathbf{A})$ & PSF simulated using modal coefficients vector $\mathbf{A}$, \\
       $\mathcal{I}(\mathcal{A})$ & cube of simulated PSFs, \\
       $\mathbf{1}_N$ & vector of ones of length $N$, \\
       $\otimes$ & Kronecker product, \\
       $\mathcal{L(\cdot)}$ & optimisation criterion / loss function, \\
       $\left\lvert \left\lvert \hspace{2pt} \cdot \hspace{2pt} \right\rvert \right\rvert_1$ & $L_1$-norm. \\
       \noalign{\smallskip}
       \hline
    \end{tabular}
\end{table}
\noindent
The entities introduced in Table~\ref{tab:DIP:notations} have the following dimensions:
\begin{align*}
    \mathbf{1}_{N} &= [1, 1, \ldots 1], & \mathbf{1}_{N} &\in \mathbb{R}^N \\
    \mathbf{A}_i &= [a_{i,0}, a_{i,1}, \ldots a_{i,M}], & \mathbf{A}_i &\in \mathbb{R}^M\,, \\
    \mathcal{A} &= [\mathbf{A}_0; \mathbf{A}_1; \ldots \mathbf{A}_N], & \mathcal{A} &\in \mathbb{R}^{N \times M}, \\
    \mathcal{I}_\mathrm{\hspace{-1pt} data} &= [\mathbf{I}_{\mathrm{\hspace{1pt} data}, \hspace{1pt} 0}; \mathbf{I}_{\mathrm{\hspace{1pt} data}, \hspace{1pt} 1}; \ldots \mathbf{I}_{\mathrm{\hspace{1pt} data}, \hspace{1pt} N}], & \mathcal{I}_\mathrm{\hspace{-1pt} data} &\in \mathbb{R}^{N \times W \times H}, \\
    \mathcal{I}(\mathcal{A}) &= [\mathbf{I}(\mathbf{A}_0); \mathbf{I}(\mathbf{A}_1); \ldots \mathbf{I}(\mathbf{A}_N)], & \mathcal{I}(\mathcal{A}) &\in \mathbb{R}^{N \times W \times H}, \\
    \mathcal{L}(x, \hat{x}) &= \frac{1}{N} \left\lvert \left\lvert \hspace{1pt} x - \hat{x} \hspace{1pt} \right\rvert \right\rvert_1. \\
\end{align*} 
\noindent
Here, $\mathcal{L}(x, \hat{x})$ is the loss function or optimisation criterion, $x, \hat{x}$ are some variables, $\mathcal{I}_\mathrm{\hspace{-1pt} data}$ represents a cube composed of PSFs captured sequentially at a sampling frequency of 10 Hz. With all the required terms defined, we are ready to introduce a few approaches to coefficient optimisation.

\vspace{10pt}
Method 1 involves averaging the recorded PSF cube $\mathcal{I}_\mathrm{\hspace{-1pt} data}$ along the temporal dimension and fitting a single modal coefficients vector to the averaged image, which is akin to the standard implementation of LIFT discussed earlier. This provides minimal computational overhead and rapid convergence due to the relatively small scale of the problem. The optimisation can then be formulated as:
\begin{displaymath}
    \hat{\mathbf{A}} = \mathrm{argmin}_{\mathbf{A}} \hspace{1pt} \mathcal{L} \left( \frac{1}{N} \hspace{-2pt} \sum_{i}^{N} \mathbf{I}_{\mathrm{\hspace{1pt} data}, \hspace{1pt} i}, \hspace{2pt} \mathbf{I}(\mathbf{A}) \right).
\end{displaymath}
\noindent
However, averaging the data cube can lead to the blurring of the fine morphological features of the PSFs and generating a residuals-driven 'halo'. It is possible to incorporate the residual-related component into the PSF model to address this issue, as demonstrated in \cite{Mugnier:08}. However, implementing this technique is beyond the scope of the current work. Therefore, we limit our discussion to PSF models driven solely by modal coefficients.
\vspace{10pt}

Method 2. In contrast to the previous method, this optimisation approach implies fitting a single coefficient vector to the PSF cube $\mathcal{I}_\mathrm{\hspace{-1pt} data}$. This forces the optimiser to find a coefficient vector $\hat{\mathbf{A}}$ that simultaneously satisfies all PSFs in the cube, thus preventing overfitting to a particular PSF sample. Given the utilisation of a single coefficient vector, the gradient updates for each coefficient in the vector are collected from all samples and then averaged. Therefore, the part of the gradients common to all samples will dominate, while the noisy component will be averaged out.
\noindent  
This estimation method can be expressed as follows, noting that the Kroeneker product $\otimes$ clones the same coefficient vector for each data sample:
\begin{displaymath}
    \hat{\mathbf{A}} = \mathrm{argmin}_{\mathbf{A}} \hspace{1pt} \mathcal{L} \left( \mathbf{1}_N \hspace{-1pt} \otimes \mathbf{I}(\mathbf{A}), \hspace{2pt} \mathcal{I}_\mathrm{\hspace{-1pt} data} \right).
\end{displaymath}
\vspace{3pt}

Method 3. Another possible approach is to independently fit a unique coefficient vector $\mathbf{A}_i$ to every individual sample $\mathbf{I}_{\mathrm{\hspace{1pt} data}, \hspace{1pt} i}$. No gradients between coefficients are shared in this case. While this approach allows for a more precise fit for each individual sample by increasing the parameter space, it also increases the likelihood of overfitting.
This approach be expressed as follows:
\begin{displaymath}
    \hat{\mathcal{A}} =
    \begin{bmatrix}
        \mathrm{argmin}_{\mathbf{A}_0} \hspace{1pt} \mathcal{L} \left( \mathbf{I}_{\mathrm{\hspace{1pt} data}, \hspace{1pt} 0}, \hspace{2pt} \mathbf{I}(\mathbf{A}_0) \right) \\
        \mathrm{argmin}_{\mathbf{A}_1} \hspace{1pt} \mathcal{L} \left( \mathbf{I}_{\mathrm{\hspace{1pt} data}, \hspace{1pt} 1}, \hspace{2pt} \mathbf{I}(\mathbf{A}_1) \right) \\
        \vdots \\
        \mathrm{argmin}_{\mathbf{A}_N} \hspace{1pt} \mathcal{L} \left( \mathbf{I}_{\mathrm{\hspace{1pt} data}, \hspace{1pt} N}, \hspace{2pt} \mathbf{I}(\mathbf{A}_N) \right)
    \end{bmatrix}.
\end{displaymath}
\noindent
The same can be written down in a slightly shorter notation as fitting a matrix of coefficients, $\hat{\mathcal{A,}}$ to the PSF cube, $\mathcal{I}_\mathrm{\hspace{-1pt} data}$.
\begin{equation}
    \hat{\mathcal{A}} = \mathrm{argmin}_{\mathcal{A}} \hspace{1pt} \mathcal{L} \left( \mathcal{I}_\mathrm{\hspace{-1pt} data}, \hspace{2pt} \mathcal{I}(\mathcal{A}) \right)\, , \quad \hat{\mathbf{A}} = \frac{1}{N} \hspace{-1pt} \sum_i^N \hat{\mathbf{A}}_i.
    \label{eq:DIP:global}
\end{equation}
\noindent
This notation is equivalent to the simultaneous estimation of the ensemble of coefficient vectors, $\mathbf{A}_i$. All coefficient vectors inside the $\mathcal{A}$ can then be averaged to get the final estimate for the quasi-static component.
\vspace{10pt}

In summary, the optimisation methods discussed above require substantial computational resources due to the increased scale of the optimisation problems. The iterative optimisation process involves repeatedly simulating hundreds of PSFs and computing Jacobians, which becomes particularly computationally demanding when using methods~2 and~3. Therefore, it is crucial to implement an efficient and high-performing code to address this issue before performing the analysis. This problem will be approached in the subsequent section.

\subsection{Introducing DIP}
\label{subsec:DIP}
Here, we present the differentiable PSF model (DIP), the \href{https://github.com/EjjeSynho/DIP}{code} designed to accelerate and facilitate extensive PSF simulations and large-scale PSF fitting tasks based on the PyTorch framework. Given it is designed for machine learning, PyTorch is well-optimised for handling massive data batches both on CPU and GPU backends. Furthermore, PyTorch offers a diverse range of optimisation algorithms that can be seamlessly integrated with DIP. Finally, DIP takes advantage of the automatic differentiation functionality of PyTorch, which allows for gradients to be computed with almost analytical precision for the PSF models of arbitrary complexity. In turn, this eliminates the necessity to derive the analytical expressions for Jacobian and Hessian matrices or to approximate them numerically with finite differences. Harnessing these advantages allows DIP to go beyond the initial mathematical formalism of LIFT, resulting in a versatile framework where the PSF model, optimiser, and loss function are distinct interchangeable components, implementing a concept elaborated upon earlier in Sect.~\ref{sec:theory}.

\subsection{Selecting the estimation approach}
\begin{figure}[h!]
   \centering
   \includegraphics[width=0.475\textwidth]{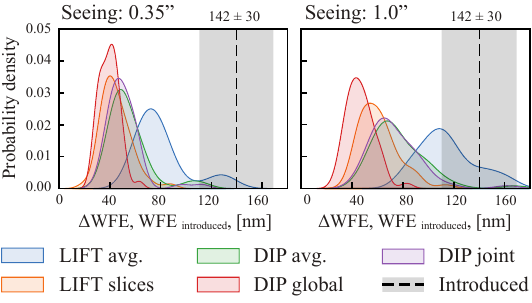}
   \caption
   {
      Comparison of different methods to retrieve the quasi-static LO modes associated with NCPAs. The 'DIP global' method shows the highest accuracy, especially in the case of larger seeing (right picture). The curves represent the distribution of RMS of the averaged delta wavefront error ($\Delta \mathrm{WFE}$). The black dashed line and grey fill show the mean RMS of introduced NCPA realisations ($\mathrm{WFE}_\mathrm{\hspace{1pt} introduced}$) with 1$\sigma$ confidence interval.
   }
   \label{fig:DIP:methods}
\end{figure}

By leveraging DIP, it becomes feasible to implement the optimisation approaches 1-3 proposed in the previous subsection. To quantitatively compare them, we conducted a test shown in Fig.~\ref{fig:DIP:methods}. The test assesses the ability of estimation methods to retrieve the quasi-static phase component associated with NCPAs accurately. The figure depicts the distribution of RMS WFE, which is computed for the difference between the temporally averaged simulated wavefront cube and the estimated static LO wavefront (see Eq. (\ref{eq:DIP:WFE}) for more details).
To do this test, a dataset of simulated PSF cubes was generated and subsequently analysed using the techniques derived from estimation methods~1-3 covered earlier in the previous section.

In Fig.~\ref{fig:DIP:methods}, 'LIFT avg.' represents the standard implementation of LIFT detailed earlier in Sect.~\ref{sec:theory}. 'DIP avg.' operates similarly to classical LIFT and implements the estimation approach 1. However, it is powered by the DIP framework, meaning it takes the same average PSF as input, while the optimisation algorithm and loss function differ. For DIP, we adopted the L-BFGS optimisation algorithm described in ~\cite{Liu:89}. As an optimisation criterion, DIP uses the Maximum Absolute Error (MAE or $L_1$-loss).
'DIP joint' implements the estimation approach 2, where a single coefficient vector is shared across all PSF samples within the cube.
'LIFT slices' refers to applying LIFT in the fashion outlined in method 3, where the fitting process is conducted on a per-sample basis.
Finally, 'DIP global' is also based on method~3. However, it simultaneously fits all coefficient vectors, essentially implementing Eq. (\ref{eq:DIP:global}).

The analysis indicates that the 'LIFT slices' and 'DIP global' methods exhibit superior precision in retrieving quasi-static LO modes compared to other listed methods. This can be explained by the fact that when the PSF model is fitted to a set of short-exposure PSFs, each PSF sample preserves the information about high spatial frequencies in its speckles, consequently facilitating the retrieval of higher-order modes. In this context, 'higher-order modes' refer to modes higher than TTF that are included in the estimation. In this test, 29 first Zernike modes were estimated. The question of choosing the number of estimated modes will be covered in the next section. Meanwhile, accurately retrieving higher orders also facilitates the accurate estimation of LO modes, as previously mentioned in Sect.~\ref{sec:onsky_1:lessons}.
In contrast, this information is lost in averaged long-exposure PSFs. In the same fashion, sharing one coefficient vector over all PSF samples in 'DIP joint' effectively acts as averaging the PSF.
It is important to note that there is a risk of overfitting when using 'LIFT slices' and 'DIP global'. However, this is counterbalanced by the final averaging across the ensemble of modal vector estimates. This is connected to the fact that PSF samples are captured sequentially in the presence of HO residuals. Therefore, all PSF samples slightly differ from each other. Thus, only the static component of corresponding estimated modal vectors remains after the averaging.

It is important to note that all DIP-based approaches demonstrated overall higher estimation accuracy compared to LIFT-based approaches. For example, 'DIP avg.' shows superior accuracy compared to classical LIFT implementation, even though the same data inputs are used in these two cases. In turn, 'DIP global', although it is essentially equivalent to 'LIFT slices', shows better accuracy for the larger seeing values. This can be attributed to utilising the L-BFGS instead of the Gauss-Newton optimisation algorithm intrinsically used in LIFT. Based on these results, we adopt the 'DIP global' as our baseline solution for NCPA calibration further in this paper.

To generate Fig.~\ref{fig:DIP:methods}, realistic simulations were performed to recreate on-sky conditions. The dataset of 100 samples was generated, where each sample features a PSF cube with $N = 300$ sequential IRLOS-like short-exposure PSFs and a corresponding cube of true phase. We selected this number of samples because there was no noticeable improvement in estimation quality when using $N > 200$. However, to ensure robustness, we opted for a slightly higher number. Each phase screen in such a cube is the sum of tip-tilt ($\mathbf{\Phi}_\text{TT}$) residuals, low-order NCPAs ($\mathbf{\Phi}_\text{LO}$), HO residual phase ($\mathbf{\Phi}_\text{HO}$), and phase diversity ($\phi_d$); $\mathbf{\Phi}_\text{LO}$ is based on the random set of the first 50 Zernike coefficients, which remain constant for each PSF within one given cube. However, every sample in the dataset features different realisations of LO coefficients. For further information on simulating the LO component, we refer to Appendix~\ref{sec:appendix:stats}. The residual phase screens are sampled from actual recorded AOF telemetry with the addition of the simulated atmospheric residue. The fast $\mathbf{\Phi}_\text{TT}$ is randomly generated for each screen based on residual TT statistics observed on IRLOS. The simulations incorporate noise and model a point source with a magnitude of $m_J = 7$ for all samples in the dataset, matching the magnitude used for the on-sky samples discussed in this paper. However, the simulations indicate that, with an exposure time of 0.1 seconds per sample, calibration targets with $m_J < 15$ are viable.
Further details on the simulation parameters can be found in Table~\ref{tab:DIP:params}.

\begin{table}[h!]
    \caption{Parameters for the calibration-stage simulations.}
    \label{tab:DIP:params}
    \centering
    \begin{tabular}{lc}
        \hline
        \noalign{\smallskip}
        Parameter & Value \\ 
        \noalign{\smallskip}
        \hline
        \noalign{\smallskip}
        Target $m_J$ & 7 \\
        Sampling time, [s] & 0.1 \\
        PSFs in a cube & 300 \\
        $a_5$ diversity, [nm] & -170 \\
        Wavelength $\lambda$, [nm] & 1600 \\ 
        Seeing, ["] & 0.35, 1.0 \\
        Introduced NCPAs, [nm RMS] & 142 $\pm$ 30 \\
        \noalign{\smallskip}
        \hline
    \end{tabular}
\end{table}

Further details on the computation of the WFE presented in Fig.~\ref{fig:DIP:methods} are provided in Eq. (\ref{eq:DIP:WFE}), along with clarifications on previously introduced notations.
Here, $\mathbf{\Phi}$ is the average of simulated wavefronts $\mathbf{\Phi}_i$; then $\mathit{\Phi}$ is the cube of simulated wavefronts; $\hat{\mathbf{\Phi}}$ is the estimated static NCPA wavefront; $\Delta \mathrm{WF}$ is the difference between introduced and retrieved averaged wavefronts $\mathrm{WF}_\mathrm{\hspace{1pt}introduced}$ and $\mathrm{WF}_\mathrm{\hspace{1pt}estimated}$; $\mathrm{WFE}(\cdot)$ is the RMS of a wavefront computed over the pupil pixels; $N_{pupil}$ is the number of valid pixels in the pupil mask, $\mathbf{P}$; the term $\mathbf{P} (\mathbf{\Phi} - \hat{\mathbf{\Phi}} )$ is summed over the pupil-plane pixels; $\odot$ is the Hadamard product; and $\mathbf{X}$ is a matrix variable.

\noindent
Thus, we have:\ 
\begin{equation}
    \begin{aligned}
        \mathbf{\Phi}_i &= \mathbf{\Phi}_\text{LO} + \mathbf{\Phi}_{i,\hspace{1pt} \text{TT}} + \mathbf{\Phi}_{i,\hspace{1pt} \text{HO}} + \phi_d , \\
        \mathit{\Phi} &= [ \mathbf{\Phi}_0; \mathbf{\Phi}_1; \ldots \mathbf{\Phi}_N ], \\
        \mathbf{\Phi} &= \frac{1}{N} \hspace{-2pt} \sum_i^N \mathbf{\Phi}_i, \\
        \hat{\mathbf{\Phi}} &= \mathbf{Z} \hat{\mathbf{A}} + \phi_d \\
        \mathrm{WFE} \left( \mathbf{X} \right) &= \sqrt{ \frac{1}{N_{pupil}} \sum \sum \left( \mathbf{P} \odot \mathbf{X} \right)^2 }, \\
        \mathrm{WF}_\mathrm{\hspace{1pt}introduced} &= \mathbf{\Phi} - \phi_d, \\
        \mathrm{WF}_\mathrm{\hspace{1pt}estimated} &= \hat{\mathbf{\Phi}} - \phi_d, \\
        \Delta \text{WF} &= \mathrm{WF}_\mathrm{\hspace{1pt} introduced} - \mathrm{WF}_\mathrm{estimated}, \\
        \text{WFE}_\mathrm{\hspace{1pt} introduced} &= \text{WFE}\hspace{2pt}(\mathbf{\Phi} - \phi_d), \\
        \Delta \mathrm{WFE} &= \mathrm{WFE}\hspace{2pt}(\Delta \mathrm{WF}). \\
    \end{aligned}
    \label{eq:DIP:WFE}
\end{equation}
\noindent

\subsection{Verifying calibration approach}
\label{subsec:DIP:verify}
\begin{figure}[H]
   \centering
   \includegraphics[width=0.5\textwidth]{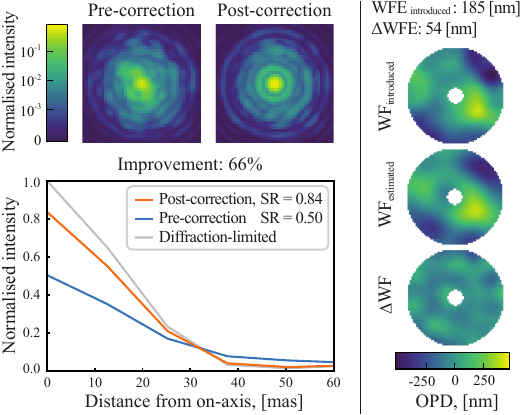}
   \caption{Averaged PSF cubes simulated with (post-correction) and without (pre-correction) compensating for the NCPAs (see left column, top row). The improvement observed in the 'post-correction' case indicates the successful retrieval of the quasi-static phase component associated with NCPAs. The right column demonstrates the corresponding wavefronts and their difference. See Eq. (\ref{eq:DIP:WFE}) for notation details. Simulated with $\lambda = 1600$ nm.}
   \label{fig:DIP:improvement}
\end{figure}

After selecting the 'DIP global' approach as the baseline method for NCPA calibration, it is now possible to illustrate its ability to correct for the quasi-static LO phase.
Figure~\ref{fig:DIP:improvement} highlights the performance improvement achieved when using 'DIP global' as the NCPA calibrator.

The top left part of the figure presents averaged PSF cubes illustrating the impact of NCPAs on a focal-plane PSF before and after applying the DIP-measured correction. The PSFs shown are simulated under identical conditions to those that produced the data for Fig.~\ref{fig:DIP:methods} (see Table~\ref{tab:DIP:params}). However, presented PSFs exclude the effect of the astigmatic diversity to improve visual clarity, although diversity was used in estimation. Specifically, the 'pre-correction' PSF cube is defined by the wavefront $\mathit{\Phi} - \mathbf{1}_N \hspace{1pt} \otimes \hspace{1pt} \phi_d$, while the 'post-correction' PSF cube was computed using the $\mathit{\Phi} - \mathbf{1}_N \hspace{1pt} \otimes \hspace{1pt} (\phi_d + \hat{\mathbf{\Phi}}$) as input wavefronts. In other words, for this plot, the correction was simulated by subtracting the estimated wavefront from the simulated wavefront cube.

The right column of this figure demonstrates the simulated and estimated averaged wavefront cubes ($\mathrm{WF}_\mathrm{\hspace{1pt} introduced}$ and $\mathrm{WF}_\mathrm{\hspace{1pt} esimated}$, respectively) associated with the depicted PSFs.
We note that 29 modes were selected for estimation due to specific SPARTA RTC limitations. As elaborated on in Sect.~\ref{sec:onsky_2:calib_verify}, compensation of NCPAs via the DSM can be implemented by adding constant modal offsets to the WFS slopes. Currently, SPARTA allows us to set up to 29 Zernike modal offsets. Although this limitation can be technically overcome, for simplicity, we standardise on $M=29$ for the entirety of this paper. Further details on selecting the optimal number of estimated modes are available in Appendix~\ref{sec:appendix:overfit}.

\subsection{Calibrated LIFT}
\label{subsec:DIP:calib_LIFT}
\begin{figure}[H]
   \centering
   \includegraphics[width=0.475\textwidth]{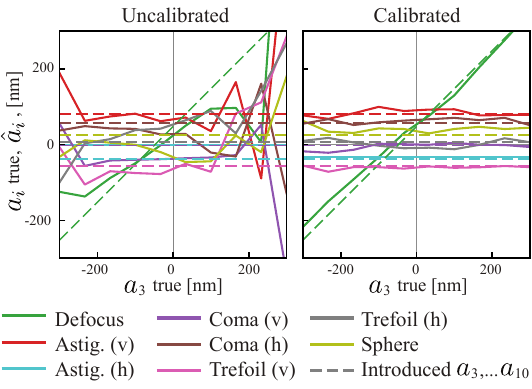}
   \caption{
   Estimation of LO modes with LIFT on a simulated dataset of PSFs with ramping defocus. This test is similar to the one shown in Fig.~\ref{fig:exp:onsky1}.
   This time, defocus was estimated for the same PSF dataset with and without introducing the pre-calibrated NCPAs prior.
   A substantial increase in LIFT estimation accuracy is observed when calibration is applied.
   Solid lines represent the estimated modal coefficients $(\hat{a}_1, \ldots \hat{a}_{10})$, dashed lines represent corresponding introduced values.
   Simulated seeing = 0.35", $m_J = 12$, $\mathrm{WFE}_\mathrm{\hspace{1pt} introduced} \approx 140$ nm RMS. See Appendix~\ref{sec:appendix:seeing} for more information on seeing selection.
   }
   \label{fig:DIP:lin_scan}
\end{figure}

After demonstrating the ability of the calibration stage to retrieve quasi-static NCPA components, we combine it with the LIFT estimation stage.
The plot in Fig.~\ref{fig:DIP:lin_scan} compares the performance of LIFT by using a calibrated prior (right) against its operation without such a prior (left).

Essentially, this figure replicates the experiment initially depicted in Fig.~\ref{fig:exp:onsky1} in a simulated environment. To produce the PSFs for the second stage, we generated a series of PSFs with varying defocus values within the range $a_3 = -300 \ldots 300$ nm RMS. The other simulation parameters were adopted from Table~\ref{tab:onsky:params_1}. In addition to PSFs for LIFT, a PSF cube was generated and preemptively estimated with DIP to retrieve the quasi-static NCPAs. In this case, simulation parameters from Table~\ref{tab:DIP:params} were utilised. Notably, the same set of NCPAs was employed for both the LIFT PSFs and the DIP PSF cube, albeit different realisations of HO and TT residuals were added in both cases.
In scenarios where calibration was applied, the prior information from DIP was incorporated into LIFT as an additional phase diversity term, effectively leading to $\phi_{d, \hspace{1pt} \text{calibrated}} = \phi_d + \hat{\mathbf{\Phi}}$.
Figure~\ref{fig:DIP:lin_scan} indicates a marked improvement in the precision of defocus estimation, as well as the estimation of other modes, upon utilising calibrated prior.

The results show that LIFT provides more accurate and stable estimates when calibrated. Most importantly, calibrations eliminate the need to estimate modes beyond TTF with LIFT since the higher orders associated with quasi-static NCPAs are 'absorbed' by the prior provided by DIP. Overall, notwithstanding their rather illustrative nature, these tests support our proposal for the two-stage approach in the simulated environment and ensure the following on-sky verification.

\section{Second on-sky tests}
\label{sec:onsky_2}

In this section, we describe how we verified the two-stage approach with a second set of on-sky tests. The organisation of this section is similar to that of the preceding one, beginning with an examination of the calibration stage in isolation, followed by an assessment of its integration with LIFT. Moreover, an additional sensitivity test complements these results. The key distinction is that all the results discussed in this section are exclusively derived from on-sky observations obtained using the VLT UT4 telescope. 

\subsection{Calibration stage on-sky}
\label{sec:onsky_2:calib_verify}
The significant challenge encountered during on-sky testing is the lack of reliable ground truth. As discussed in Sect.~\ref{sec:onsky_1}, controlling the defocus introduced by shifting the focal stage of MUSE is feasible. However, it is hard to characterise the NCPAs met in the IRLOS optical path independently with another method. Furthermore, NCPAs can be induced not only by phase errors but also by amplitude distortions. However, we assume that all NCPA-related effects observed in the recorded on-sky PSFs are exclusively associated with phase.

\begin{figure}[h!]
   \centering
   \includegraphics[width=0.475\textwidth]{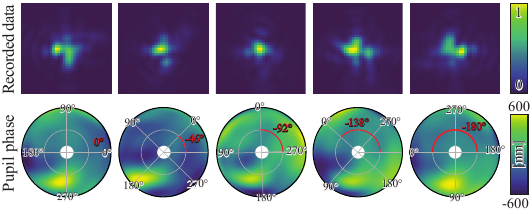}
   \caption{
   Wavefronts (bottom row) estimated using on-sky PSFs (top row) with different manually set field derotator angles. A notable 'phase bump' can be consistently seen across all estimates. Note that the images of wavefronts are rotated in accordance with the derotator angle. The samples were recorded under the seeing $\approx 1.2$", $\lambda = 1600 \pm 20$ nm.
   }
   \label{fig:exp:onsky2_derotator}
\end{figure}

Due to the absence of reliable ground truth, we first conducted qualitative tests as presented in Fig.~\ref{fig:exp:onsky2_derotator}. This involved recording a series of measurements using IRLOS. The procedure involved manually rotating the MUSE field derotator and recording a PSF cube for each angle value. Here, the consistency of the estimated phase over different derotator angles serves as a proxy for veracity. The parameters of the acquisition process are detailed in Table~\ref{tab:onsky:params_2_derotator}. It is important to remember that the optical design of IRLOS does not support the simultaneous use of a narrow band filter and the astigmatic lens. Consequently, to facilitate data collection in a monochromatic regime, we introduced astigmatic diversity using the deformable secondary mirror (DSM) of UT4 instead of the lens.

It is worth noting that using a narrow-band filter can introduce NCPAs (for example, those encountered when placing a glass plate in a convergent beam). These aberrations will then be present during the calibration but absent during the focus retrieval stage, potentially leading to biased estimations. To alleviate this issue, it is possible to perform NCPA calibration using multiple narrow-band filters and compare the results. However, in these experiments, we assume that the narrow-band filter is free of aberrations.

\begin{table}
    \caption{Acquisition parameters for the derotator test.}
    \label{tab:onsky:params_2_derotator}
    \centering
    \begin{tabular}{lc}
        \hline
        \noalign{\smallskip}
        Parameter & Value \\ 
        \noalign{\smallskip}
        \hline
        \noalign{\smallskip}
        Target $m_J$ & 8 \\
        Exposure time per sample, [s] & 0.1 \\
        PSFs per cube $N$ & 600 \\
        Number of introduced angles & 9 \\
        Angles range, [deg] & 0, 23, \ldots, 161, 180 \\
        $a_5$ diversity (DSM), [nm] & 170 \\
        Wavelength $\lambda \pm \Delta \lambda$, [nm] & 1600 $\pm$ 20 \\ 
        \noalign{\smallskip}
        \hline
    \end{tabular}
\end{table}

The PSF cubes shown in Fig.~\ref{fig:exp:onsky2_derotator} underwent estimation using the 'DIP global' calibration method. The retrieved phase screens depicted in Fig.~\ref{fig:exp:onsky2_derotator}, demonstrate the consistency across various derotator angles. The observed variability in the measurements can be attributed to imprecise estimations and the existence of NCPAs that remain non-rotating due to their origin beyond the derotator in the optical path. Notably, a distinct bump-like feature is consistently observed as shifting around the field with different derotator angles. Even though our method is agnostic to the origin of this anomaly, it still offers a means to quantify it.

\begin{figure}[h!]
   \centering
   \includegraphics[width=0.475\textwidth]{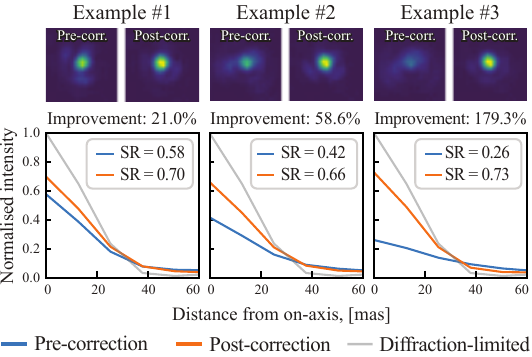}
   \caption{
   Improvement in on-sky PSF quality after correcting NCPAs with DSM. The top row shows the recorded in-focus PSFs, while the corresponding radial profiles are displayed in the bottom row. For visual clarity, the astigmatic diversity used during estimation was removed while recording these PSFs. Seeing is $\approx 0.5$", $\lambda = 1600 \pm 20$ nm.
   }
   \label{fig:exp:onsky2_improve}
\end{figure}

While primarily qualitative, this experiment demonstrates the capacity to retrieve relatively consistent phase estimates using DIP. Further tests were conducted to examine if DIP-measured NCPAs could be compensated for by using DSM to improve the quality of the in-focus IRLOS PSFs. The results are illustrated in Fig.~\ref{fig:exp:onsky2_improve}.

Here, the focal-plane PSFs captured before and after correction are displayed. The idea behind this figure is similar to Fig.~\ref{fig:DIP:improvement}, but with an on-sky execution. The phase retrieved by DIP was sent to DSM as modal offsets to perform the correction.
The leftmost result was obtained when DIP was applied to the system 'as is' and without introducing artificial errors. In this case, IRLOS was optimally refocused beforehand to achieve the best PSF quality. The middle and the right plots represent the cases affected by some uncontrolled defocus bias, which was accurately measured by DIP and subsequently corrected using DSM commands. In the same fashion as in Fig.~\ref{fig:DIP:improvement}, astigmatic diversity was employed during the data acquisition used for DIP estimation, but was not present in the optical path while recording these PSFs to enhance visual clarity.

In summary, these findings validate the ability of DIP to accurately retrieve meaningful information about the quasi-static phase component in the complex on-sky environment, further supporting the hypothesis that some measured NCPAs may indeed relate to phase aberrations, at least to some extent.
The forthcoming investigation will involve integrating the calibration process with LIFT for on-sky testing.

\subsection{Two-stage approach on-sky}
\label{sec:onsky_2:calibrated}

An experiment presented in this subsection is very similar to the one explained earlier in Sect.~\ref{sec:onsky_1}. However, this time, the PSF cube for NCPA estimations was recorded before recording the dataset of defocused PSFs for LIFT, which is the key difference. Another difference is the monochromatic nature of the recorded LIFT PSFs, which were captured using the H-band filter and with diversity introduced by DSM. This deviation was necessitated by the limited on-sky time available. Nevertheless, these changes are minor and do not alter the fundamental conditions of the experiment. The acquisition parameters for this test are listed in Table~\ref{tab:onsky:params_2_NCPAs}.
It is crucial to highlight that for this particular test, the NCPAs identified through DIP were not physically corrected using the DSM, unlike in the experiment conducted in the previous subsection. Instead, these measurements were added to the phase diversity term inside LIFT. This approach aligns with the procedures discussed earlier in the simulated experiment shown in Fig.~\ref{fig:DIP:lin_scan}. Although it was technically feasible to pre-compensate for NCPAs using DSM and record an additional dataset of corrected PSFs with ramping defocus, it would require more on-sky time.

\begin{table}
    \caption{Acquisition parameters for NCPAs retrieval test.}
    \label{tab:onsky:params_2_NCPAs}
    \centering
    \begin{tabular}{lc}
        \hline
        \noalign{\smallskip}
        \textbf{Parameter} & \textbf{Value} \\ 
        \noalign{\smallskip}
        \hline
        \noalign{\smallskip}
        Target $m_J$ & 7.1 \\
        Exposure time per sample, [s] & 0.1 \\
        PSFs per cube $N$ & 300 \\
        $a_5$ diversity (DSM), [nm] & 170 \\
        Wavelength $\lambda \pm \Delta \lambda$, [nm] & 1600 $\pm$ 20 \\ 
        \noalign{\smallskip}
        \hline
    \end{tabular}
\end{table}

\begin{figure}
   \centering
   \includegraphics[width=0.475\textwidth]{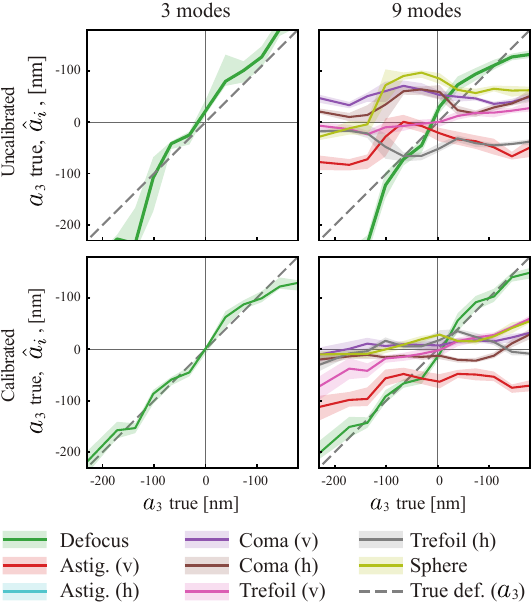}
   \caption{
   LIFT estimates for the second set of defocused on-sky PSFs. The experiment is similar to that shown in Fig.~\ref{fig:exp:onsky1}, but this time, pre-calibration was also performed, resulting in higher estimation accuracy for defocus and higher-order modes (bottom row) compared to the uncalibrated case (upper row). The error bars are the 1$\sigma$ confidence interval. NGS is $m_J = 7.1$, seeing is $\approx 0.5$".
   }
   \label{fig:exp:onsky2_lin}
\end{figure}

\subsection{Two-stage approach in the low-flux conditions}
\label{sec:onsky_2:sensitive}

\begin{figure}[h!]
   \centering
   \includegraphics[width=0.475\textwidth]{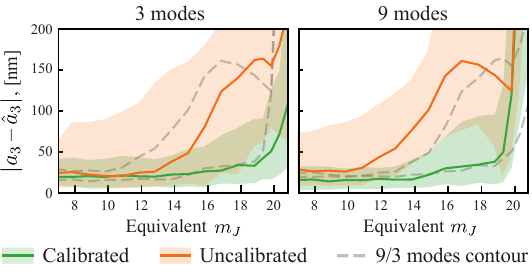}
   \caption{Sensitivity scans for the on-sky PSFs dataset. Synthetic noise was added to simulate ranging flux levels. The results again demonstrated a notable improvement in accuracy and stability defocus retrieval when NCPA pre-calibration (green line) compared to the direct estimation with LIFT.}
   \label{fig:exp:onsky2_sensitivity}
\end{figure}

It is important to note that all the results previously detailed in this paper were conducted in high-S/N conditions. However, as mentioned earlier, one of the main motivations for using LIFT is its robustness to high noise. Hence, we conducted an additional sensitivity test demonstrated in Fig.~\ref{fig:exp:onsky2_sensitivity}.
This plot presents the absolute error of defocus computed for each sample as $\mid a_3 - \hat{a}_3 \mid$, where $a_3, \hat{a}_3$ are introduced and estimated defocus values. The introduced defocus spans from -170 to 170 nm RMS. 

This test uses the same dataset as in Figure~\ref{fig:exp:onsky2_lin}. Essentially, it is a defocus ramp estimation akin to the one from the previous subsection. However, here, it is performed under varying S/Ns. The PSFs are re-normalised to different flux levels with the addition of simulated noise to achieve the desired S/N. The semi-synthetic noisy PSFs are subsequently estimated with LIFT, both with and without employing the calibrated prior. Both three (TTF) and nine modes $( a_1, \ldots a_{10} )\setminus \{ a_5 \}$ are estimated. Ten semi-synthetic PSFs with different noise realisations are generated for each magnitude and defocus value. The flux is normalised to 150000 photons per UT4 aperture per second for $m_J = 15$. An equivalent $m_J$ is obtained by analysing the photons collected in the H-band, namely, the band in which PSFs were originally recorded. The acquisition parameters of the original on-sky PSFs are consistent with the ones listed in Table~\ref{tab:onsky:params_2_NCPAs}.

Figure~\ref{fig:exp:onsky2_sensitivity} illustrates that, in practice, employing calibration significantly enhances the accuracy and stability of LIFT estimates, particularly under low-flux conditions.
It also confirms that when calibration is applied, it is sufficient to retrieve only TTF modes without sacrificing the quality of the defocus measurements due to model incompleteness, which aligns with findings from earlier simulated experiments. Moreover, the TTF-only (three modes) estimation fails at slightly higher magnitudes compared to the case with additional modes estimated (nine modes).
These results reinforce the efficacy of incorporating a calibration strategy and support the core promise of this paper, demonstrating its validity even under the challenges met in the real system in realistic on-sky conditions.

\section{Conclusion}
\label{sec:conclusion}

In this study, we conducted one of the first on-sky validations of the LIFT technique, exploring its application challenges in real-world settings using IRLOS, the low-order WFS of MUSE-NFM. Our findings show that LIFT can successfully operate as a LO WFS on the real system. However, our investigation also highlights that its accurate and consistent operation necessitates estimating additional modes beyond defocus. This requirement stems from the complex and polychromatic nature of real on-sky PSFs influenced by the strong uncompensated NCPAs. To circumvent these challenges and enhance the accuracy and stability of  LIFT, we proposed a two-stage strategy, where the estimation of additional NCPA modes is offloaded to a distinct (offline) calibration step that provides a prior that is later utilised by LIFT during the (online) focus retrieval stage. Both the simulated and experimental verification of this approach demonstrates that adopting this method improves the accuracy of LIFT, including the low-flux scenario.
In summary, these findings indicate that LIFT is a very flexible and capable solution for LO WFSing. It can be effectively implemented in real systems and can operate with suitable precision when it is properly pre-calibrated. Furthermore, the calibration technique itself can be utilised as a minimally invasive method for direct on-sky NCPA characterisation.
Most importantly, while our study focuses on IRLOS, the methodologies discussed here are adaptable and potentially beneficial for a wide range of focal-plane imaging systems beyond the specific context of IRLOS.
To date, applying LIFT within a pre-calibrated framework has not been tested using the astigmatic lens. Although the experiments described in Sect.~\ref{sec:onsky_2:sensitive} largely reflect the operational conditions anticipated for LIFT, validating the method under conditions close to the real operations remains crucial. Therefore, more on-sky tests are necessary.
In addition, this work primarily studies LIFT in a 'bootstrapping' regime, where large defocus values are retrieved. However, closing the slow defocus loop on LIFT and testing it in this operational regime remains a topic for future research.
Moreover, the issue of the temporal stability of NCPAs requires thorough characterisation and rigorous consideration in future studies.

The prospect of integrating DIP as an alternative to LIFT in the focus retrieval stage must also be considered. The present adoption of LIFT can be attributed to its relative simplicity and reduced computational requirements in contrast to the more resource-intensive implementation of DIP.
Moreover, ongoing efforts to apply LIFT and DIP within the context of petaling and LWE sensing, as initiated in \citet{Agapito:22}, present a promising avenue for expanding this research focus.

\begin{acknowledgements}

We would like to thank Johann Kolb and Pavel Shchekaturov for their assistance in preparing and conducting the on-sky experiments.

This work benefited from the support of the European Southern Observatory, French National Research Agency (ANR) with \emph{WOLF (ANR-18-CE31-0018)}, \emph{APPLY (ANR-19-CE31-0011)} and \emph{LabEx FOCUS (ANR-11-LABX-0013)}; the Programme Investissement Avenir \emph{F-CELT (ANR-21-ESRE-0008)}, the \emph{Action Sp\'ecifique Haute R\'esolution Angulaire (ASHRA)} of CNRS/INSU co-funded by CNES, the \emph{ECOS-CONYCIT} France-Chile cooperation (\emph{C20E02}), the \emph{ORP-H2020} Framework Programme of the European Commission’s (Grant number \emph{101004719}), \emph{STIC AmSud (21-STIC-09)}, the R\'egion Sud and the french government under the \emph{France 2030 investment plan}, as part of the \emph{Initiative d'Excellence d'Aix-Marseille Universit\'e A*MIDEX, program number AMX-22-RE-AB-151}. 
\end{acknowledgements}

\bibliographystyle{aa}
\bibliography{aanda}

\begin{appendix}

\section{Statistics of LO modes}
\label{sec:appendix:stats}

\begin{figure}[h!]
   \centering
   \includegraphics[width=0.475\textwidth]{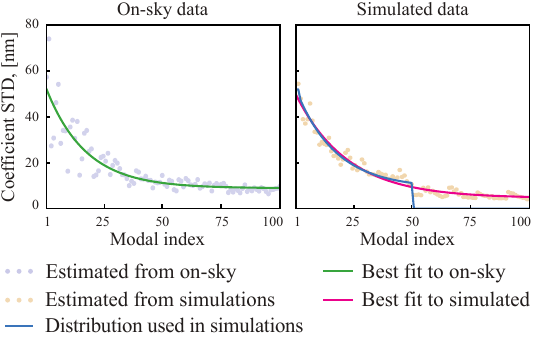}
   \caption{Standard deviation of the LO coefficients obtained using on-sky PSFs (left) and realistically simulated PSFs (right). The synthetic PSFs were generated using the LO distribution law (blue line), derived from the distribution previously estimated  from on-sky data (green line).}
   \label{fig:append:num_modes}
\end{figure}

In the left plot on Fig.~\ref{fig:append:num_modes}, the standard deviation (STD) of low-order (LO) modes was calculated using the unbiased 'DIP global' method on a dataset of real short-exposure on-sky PSFs (indicated by pale blue dots), where 100 Zernike modes were estimated. An empirical exponential law was then fitted to this data, shown by the green line.

After 50-60 modes (left plot), the STD of coefficients reaches a plateau. We attribute this plateau primarily to estimation noise due to cross-talk with HO residuals rather than to NCPAs. Consequently, in our simulations, we have limited the number of LO modes modelled to 50.

The right plot on Fig.~\ref{fig:append:num_modes} shows the STD of modes (represented by orange dots) estimated from synthetically generated PSFs, for which 100 Zernike modes were again estimated. These PSFs were created based on the law (blue line) that derives from the best fit to the on-sky data (green line). However, it was truncated to include only 50 modes, and the standard deviations of the tip and tilt were equalised.

The magenta line represents the best fit for the synthetic distribution of coefficients STD. From the distribution of the estimated coefficients, we can see that although only 50 modes were introduced into the simulated LO, the estimates also exhibit plateauing behaviour. Similarly to on-sky, it can also be attributed to the coupling with HO residuals, which are simulated, pointing out that the truncation approach is valid.

The magenta line illustrates the best fit for the synthetic distribution of coefficient STDs. Observations from this distribution reveal that, although only 50 modes were incorporated into the simulated LO, the estimates still display a plateauing trend for higher orders. This is similar to the on-sky results and is likely due to the coupling with HO residuals, indirectly confirming the validity of the truncation approach.

\section{Modal cross-talk and overfitting}
\label{sec:appendix:overfit}

\begin{figure}
   \centering
   \includegraphics[width=0.475\textwidth]{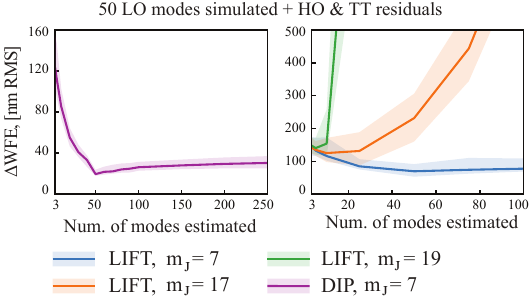}
   \caption{Impact of the number of estimated modes on estimation error. The results are shown for 'DIP global' (left) and 'LIFT avg.' (right) methods. The right plot demonstrates that reducing the number of estimated modes enhances accuracy in low-flux scenarios.}
   \label{fig:append:crosstalk}
\end{figure}

To illustrate the modal cross-talk (coupling) issue, we conducted a set of tests (see Fig.~\ref{fig:append:crosstalk}) in which the PSF dataset was simulated with a fixed number of introduced LO modes. Meanwhile, the number of estimated modes varied.

For the DIP (left plot), the optimal number of estimated modes equals the number of modes introduced. Below 50 modes, the model is not representative enough, leading to high estimation error. In contrast, sensing more than 50 modes causes overfitting, which leads to the modal cross-talk that causes deviations from the true values.

These results can be related to the practical question of determining the optimal number of modes for calibration. The findings illustrated in the left plot of Fig.~\ref{fig:append:num_modes} suggest that the most pertinent information from LO is contained within the first 50-60 orders. Thus, this number of estimated modes can be considered optimal for the calibration stage, although in practice, it was limited to 29 to ease the hardware implementation.

Figure~\ref{fig:append:crosstalk} shows the relationship between the estimation error of LIFT and the number of estimated modes for different NGS magnitudes. For bright targets, the highest accuracy is achieved with 50 modes. However, with a rising target magnitude, including fewer modes helps to regularise the problem better, prevents coupling, and reduces noise propagation.

\section{Seeing selection}
\label{sec:appendix:seeing}

\begin{figure}
   \centering
   \includegraphics[width=0.475\textwidth]{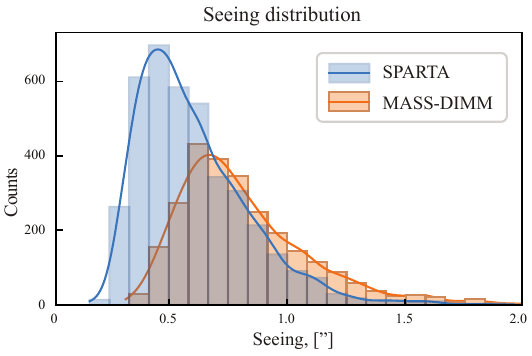}
   \caption{Seeing distribution at Paranal from 2015 to 2020 as measured by MASS-DIMM and SPARTA.}
   \label{fig:append:seeing_dist}
\end{figure}

The simulation depicted in Fig.~\ref{fig:append:seeing_dist} illustrates the distribution of seeing measured by SPARTA, based on WFS data from AOF, and by MASS-DIMM. Both methods independently estimate the seeing, although a noticeable shift is observed, with MASS-DIMM consistently being more pessimistic. The same results were earlier presented in \citet{Fetick:19}.

It is crucial to highlight that the seeing values presented in this paper with on-sky data are measured using MASS-DIMM. In contrast, the simulated seeing values are more related to SPARTA. This is because the simulated TT and HH residuals were normalised based on their expected values, which were mapped to the measurements taken by SPARTA. In this case, a seeing value of 0.35" indicates quite good conditions, whereas 1" suggests rather poor conditions. This paper utilises these two values in simulations to reflect both cases.

\end{appendix}

\end{document}